\documentclass[amsmath, aps, pra, twocolumn, floatfix]{revtex4-1}
\usepackage{graphicx}

\begin{document} 
	
\title{Repulsive Casimir-Polder potentials of low-lying excited states of a multilevel alkali-metal atom near an optical nanofiber}
	
\author{Fam Le Kien,$^{1}$ D. F. Kornovan,$^{2}$  
	S\'{i}le Nic Chormaic,$^{3}$ and Thomas Busch$^{1}$}

\affiliation{$^1$Quantum Systems Unit, Okinawa Institute of Science and Technology Graduate University, Onna, Okinawa 904-0495, Japan\\
	$^2$ ITMO University, Birzhevaya liniya 14, 199034 St. Petersburg, Russia\\
	$^3$Light-Matter Interactions Unit, Okinawa Institute of Science and Technology Graduate University, Onna, Okinawa 904-0495, Japan
}
	
\date{\today}
	
\begin{abstract}
	
We study the Casimir-Polder potential of a multilevel alkali-metal atom near an optical nanofiber. We calculate the mean potential of the atom in a fine-structure level. 
We perform numerical calculations for the Casimir-Polder potentials of the ground state and a few low-lying excited states of a rubidium atom.  We show that, unlike the potential of the ground state, which is negative and attractive, the potential of a low-lying excited state may take positive values, oscillate around the zero value with a decaying amplitude, and become repulsive in some regions of atom-to-surface distances.
We observe that, for a nanofiber with a radius of 200 nm, the potential for the state $8S_{1/2}$ of a rubidium atom achieves a  positive peak value of about 17 $\mu$K at a distance of about 150 nm from the fiber surface, and becomes rather strongly repulsive in the region of distances from 150 to 400 nm. 
We also calculate the nanofiber-induced  shifts of the transition frequencies  of the atomic rubidium $D_2$ and $D_1$ lines. We find that the shifts are negative in the region of short distances, become positive, and oscillate around the zero value with a decaying amplitude in the region of large distances.

\end{abstract}
	
\maketitle

\section{Introduction}

Over the last two decades, studies on the interaction between atoms and guided light of optical nanofibers have received a lot of interests \cite{TongNat03,review2016,review2017,Nayak2018}. Optical nanofibers are ultrathin tapered fibers that have a subwavelength diameter and significantly differing core and cladding refractive indices \cite{TongNat03}. They allow guided light, tightly confined radially, to propagate along the fiber for a long distance (with several millimeters being typical) and to interact efficiently with nearby quantum or classical emitters, absorbers, and scatterers \cite{review2016,review2017,Nayak2018}. 
Optical nanofibers have been used in various applications including sensing \cite{sensing1,sensing2},
nonlinear optics \cite{nonlinear1,nonlinear2}, 
quantum optics \cite{quantumoptics}, 
particle manipulation \cite{particle1,particle2},
and as optical couplers in photonics \cite{coupler1,coupler2}.
Such nanofibers have recently become an important tool in atomic physics and have been used for trapping cold atoms \cite{onecolor,twocolor,Vetsch2010,Goban2012,tftrap},
efficient channeling of emission from atoms into guided modes \cite{cesiumdecay,Nayak2007,Nayak2008}, efficient absorption of guided light by atoms \cite{absorption,Sague2007},
collective excitations of atoms \cite{collective}, 
collective strong coupling of atoms to a ring cavity \cite{ring},
generation of Rydberg states of atoms near a dielectric surface \cite{Rajasree2020}, excitation of quadrupole transitions of atoms \cite{quadrupole,Ray2020}, and atomic lifetime measurements \cite{lifetime}.

The presence of a macroscopic body in the space surrounding an atom modifies the spontaneous emission of the atom, shifts the energy levels of the atom, and exerts a position-dependent force on the atom \cite{nano-optics,Buhmann2012}. Such effects have been studied for a large number of systems \cite{nano-optics,Buhmann2012}. In the particular case of an atom near a nanofiber, the modifications of the radiative decay have been investigated in the context of a two-level atom \cite{Jhe,Tromborg,Klimov} as well as a realistic multilevel alkali-metal atom \cite{cesiumdecay,Fam2008,sponhigh,Solano2019,Stourm2020}. The shift in the energy level of the atom caused by the presence of the body depends on the position of the atom and is known as the Casimir-Polder potential. In the nonretardation regime, the body-induced  potential is sometimes referred to as the van der Waals potential \cite{Buhmann2012}. The nonretarded van der Waals potential of an atom or a molecule in the vicinity of a dielectric or metallic cylinder has been calculated using a complete set of eigenmodes of the electric scalar potential and the propagator method \cite{Boustimi2002,Boustimi2003}.  Using the electrostatic approximation and the image-charge formalism, a different study has been developed for the van der Waals interaction between an atom and the convex surface of a nanocylinder \cite{Frawley2012}. Based on the powerful Green tensor technique and the exact center-of-mass equation of motion, a systematic theory has been developed for the Casimir-Polder force on a multilevel atom near a dispersing and absorbing magnetodielectric body \cite{Buhmann2004}. With the use of the eigenmode function technique and the Hamiltonian formalism, the force of light on a two-level atom near a nanofiber has been investigated \cite{chiralforce,chiralforcefull}.

Recently, the Casimir-Polder potential of a rubidium atom in a Rydberg state near a nanofiber has been calculated \cite{Stourm2020} using the nonretardation approximation \cite{Ellingsen2011}. In the numerical calculations of Ref.~\cite{Stourm2020}, the multilevel structure of atomic rubidium has been accounted for  by summing up the contributions of the individual transitions between a large number of magnetic sublevels. Several related works have been reported \cite{Ellingsen2010,Block2017,Safari2008,HTDung2016,vdW}.  The Casimir-Polder potential of a Rydberg-state rubidium atom in a metallic cylindrical cavity  has been studied \cite{Ellingsen2010}. The van der Waals interaction potentials between Rydberg atoms near flat perfect mirrors \cite{Block2017}, between two-level atoms near magnetoelectric spheres \cite{Safari2008}, and between two-level atoms inside a hollow cylindrical waveguide \cite{HTDung2016} or outside a solid cylindrical waveguide \cite{HTDung2016,vdW} have been calculated.

In this paper, we study the Casimir-Polder potential of a multilevel alkali-metal atom near an optical nanofiber  without relying on the nonretardation approximation. We calculate the mean potential of the atom in a fine-structure level. We use the sum rules for $3j$ and $6j$ symbols to carry out analytically the summations over the relevant sublevels. We perform numerical calculations for the Casimir-Polder potentials of the ground state and a few low-lying excited states of a rubidium atom. We show that, due to the retardation and the interference between the emitted and reflected waves, the Casimir-Polder potential of a low-lying excited state may take positive values, oscillate around the zero value with a decaying amplitude, and become repulsive in some regions of atom-to-surface distances.

The paper is organized as follows. In Sec.~\ref{sec:theory} we derive an analytical expression for the Casimir-Polder potential of a multilevel alkali-metal atom  near a macroscopic body. In Sec.~\ref{sec:num}, we present the results of numerical calculations for a rubidium atom near a nanofiber. Finally, we conclude in Sec.~\ref{sec:summary}.

\section{Theory}
\label{sec:theory}

In this section, we study the Casimir-Polder potential of an atom near an arbitrary nonmagnetic dielectric or metallic body [see Fig.~\ref{fig1}(a) for the specific case of a nanofiber]. We use the theory of Ref.~\cite{Buhmann2004}, which was developed for a multilevel atom, to calculate the mean potential of an alkali-metal atom in a fine-structure state, and
use the sum rules for $3j$ and $6j$ symbols to carry out analytically the summations over the relevant sublevels.

\subsection{Casimir-Polder potential of a two-level atom}

\begin{figure}[tbh]
	\begin{center}
		\includegraphics{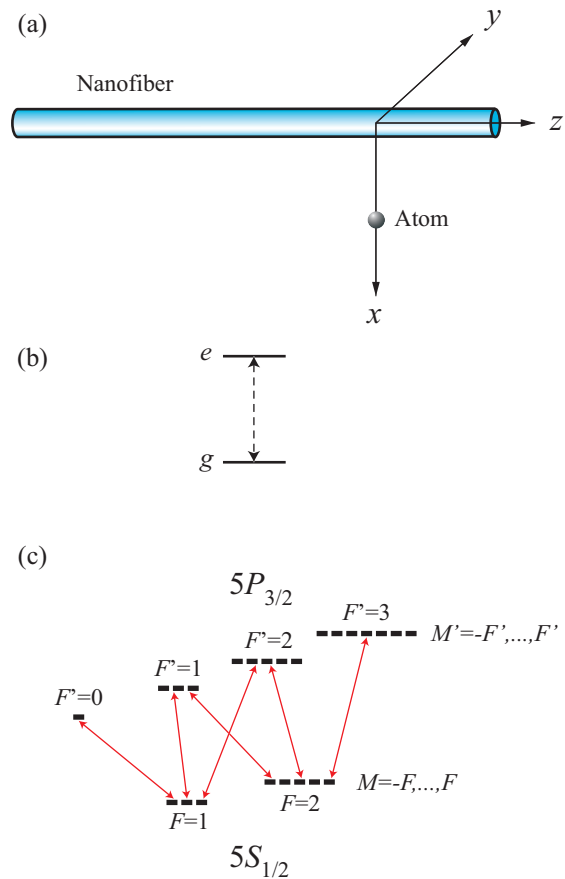}
	\end{center}
	\caption{ (a) Atom in the vicinity of an optical nanofiber. (b) Schematic of the energy levels of a two-level atom.
		(c). Schematic of the hyperfine-structure levels of the states 5$P_{3/2}$ and 5$S_{1/2}$ of a rubidium-87 atom.
	}
	\label{fig1}
\end{figure}

We start by considering the model of a two-level atom near a macroscopic body. 
The atom has an upper energy level $|e\rangle$ and a lower energy level $|g\rangle$ [see Fig.~\ref{fig1}(b)], with energies $\hbar\omega_e$ and $\hbar\omega_g$, respectively. The atomic transition frequency is $\omega_0=\omega_e-\omega_g$. 
The electric dipole transition between the levels $|e\rangle$ and $|g\rangle$ is allowed.
We use Cartesian coordinates $\{x,y,z\}$. 

The Casimir-Polder potential of the atom initially prepared in an eigenstate near a body is the body-induced part of the Lamb shift of the energy level \cite{Buhmann2012}.
In the framework of the perturbation theory,
the Casimir-Polder potentials $U_g$ and $U_e$ of the atom in the ground and excited states, respectively, are given as \cite{Buhmann2004}
\begin{eqnarray}\label{a1}
	U_g&=&\frac{\omega_0}{\pi\epsilon_0 c^2}
	\int\limits_0^{\infty}d u\frac{u^2}{\omega_0^2+u^2}
	\mathbf{d}\cdot\mathbf{G}^{(\mathrm{sc})}(\mathbf{R},\mathbf{R};i u)\cdot\mathbf{d}^*,\nonumber\\
	U_e&=&-\frac{\omega_0}{\pi\epsilon_0 c^2}
	\int\limits_0^{\infty}d u\frac{u^2}{\omega_0^2+u^2}
	\mathbf{d}\cdot\mathbf{G}^{(\mathrm{sc})}(\mathbf{R},\mathbf{R};i u)\cdot\mathbf{d}^*\nonumber\\
	&&\mbox{}-\frac{\omega_0^2}{\epsilon_0 c^2}\mathbf{d}\cdot\mathrm{Re}[\mathbf{G}^{(\mathrm{sc})}(\mathbf{R},\mathbf{R};\omega_0)]\cdot\mathbf{d}^*.
\end{eqnarray}
Here, $\mathbf{d}$ is the dipole matrix element, $\mathbf{R}$ is the position of the atom outside the body,  and $\mathbf{G}^{(\mathrm{sc})}$ is the scattering part of the Green tensor of the medium in the presence of the body.
We note that $\mathrm{Im}[\mathbf{G}^{(\mathrm{sc})}(\mathbf{R},\mathbf{R};i u)]=0$, that is, 
$\mathbf{G}^{(\mathrm{sc})}(\mathbf{R},\mathbf{R};i u)=\mathrm{Re}[\mathbf{G}^{(\mathrm{sc})}(\mathbf{R},\mathbf{R};i u)]$ \cite{Buhmann2012}. Equations (\ref{a1}) are valid for not only the nonretardation regime, where
$\Delta R\ll\lambda_0$, but also the retardation regime, where $\Delta R \gtrsim\lambda_0$. Here, $\Delta R$ is the atom-to-body distance and $\lambda_0=2\pi c/\omega_0$ is the wavelength of the atomic transition.
The Casimir-Polder potentials in the nonretardation regime are sometimes referred to as the van der Waals potential \cite{Buhmann2012}. 

The first and second terms in the expression for $U_e$ in Eqs.~(\ref{a1}) are respectively the nonresonant part $U_e^{(\mathrm{nres})}$ and the resonant part $U_e^{(\mathrm{res})}$ of the Casimir-Polder potential of the excited state $|e\rangle$.
Thus, we can write  \cite{Buhmann2004} 
\begin{equation}\label{a2}
	U_e=U_e^{(\mathrm{nres})}+U_e^{(\mathrm{res})},
\end{equation} 
where
\begin{eqnarray}\label{a3}
	U_e^{(\mathrm{nres})}
	&=&-\frac{\omega_0}{\pi\epsilon_0 c^2}
	\int\limits_0^{\infty}d u\frac{u^2}{\omega_0^2+u^2}
	\mathbf{d}\cdot\mathbf{G}^{(\mathrm{sc})}(\mathbf{R},\mathbf{R};i u)\cdot\mathbf{d}^*,\nonumber\\
	U_e^{(\mathrm{res})}&=&-\frac{\omega_0^2}{\epsilon_0 c^2}\mathbf{d}\cdot\mathrm{Re}[\mathbf{G}^{(\mathrm{sc})}(\mathbf{R},\mathbf{R};\omega_0)]\cdot\mathbf{d}^*.
\end{eqnarray}

The potential $U_g$ for the ground state $|g\rangle$ does not contain a resonant part. In other words, $U_g$ is nonresonant. It is clear that $U_g$ is opposite to the nonresonant part $U_e^{(\mathrm{nres})}$ of $U_e$, that is, $U_g=-U_e^{(\mathrm{nres})}$.

For a two-level atom with an isotropically averaged dipole, we must formally replace the factor $\mathbf{d}\cdot\mathbf{G}^{(\mathrm{sc})}\cdot\mathbf{d}^*$ in Eqs.~(\ref{a1}) by the factor $|d|^2\mathrm{Tr}\,(\mathbf{G}^{(\mathrm{sc})})/3$. It is clear that the expressions for the ground-state potential $U_g$ and the excited-state potential $U_e$ of the isotropic two-level atom contain the same effective coupling factor $|d|^2\mathrm{Tr}\,(\mathbf{G}^{(\mathrm{sc})})/3$, unlike the case of an atom with two energy level terms [see Eqs.~(\ref{a9}) in the next subsection]. 

\subsection{Casimir-Polder potential of a multilevel alkali-metal atom}

We now consider a multilevel alkali-metal atom [see Fig.~\ref{fig1}(c)]. We calculate the mean Casimir-Polder potential of the atom initially prepared in a hfs level of a fine-structure state. The energy levels of the atom are specified in an arbitrary Cartesian coordinate system $\{x_Q,y_Q,z_Q\}$, where $z_Q$ is the direction of the quantization axis. We assume that the surface-induced mixing between different hfs levels and between different fine-structure levels is negligible. We take into account the contributions of electric dipole transitions but neglect the contributions of electric quadrupole transitions. We note that this approximation is valid when the atom is not highly excited and the atom-to-body distance is not too small \cite{Stourm2020}.
 
We first examine the contributions of the transitions between the sublevels of an upper fine-structure level $|n'J'\rangle$ and a lower fine-structure level $|nJ\rangle$, where $J'$ and $J$ are the total electronic angular momenta, and $n'$ and $n$ are the sets of remaining relevant quantum numbers (such as the principal quantum number, the electron orbital angular momentum, the electron spin, and the nuclear spin). We temporally neglect the contributions of other fine-structure levels to the potentials of these two levels $|n'J'\rangle$ and $|nJ\rangle$.

Let $|e\rangle\equiv|n'J'F'M'\rangle$ and $|g\rangle\equiv|nJFM\rangle$ be the magnetic sublevels
of the hfs levels $|n'J'F'\rangle$ and $|nJF\rangle$ of the upper and lower fine-structure levels $|n'J'\rangle$ and $|nJ\rangle$, respectively. Here, $F'$ and $F$ are the total atomic angular momenta and $M'$ and $M$ are the corresponding magnetic quantum numbers. For illustration, we show in Fig.~\ref{fig1}(c) the hfs levels and the magnetic sublevels of 
the ground state $5S_{1/2}$ and the excited state $5P_{3/2}$ of a rubidium-87 atom \cite{coolingbook}.

We denote the energies of the upper and lower levels as $\hbar\omega_{e}$ and $\hbar\omega_{g}$, respectively. In the absence of the magnetic field,   $\hbar\omega_{e}$ and $\hbar\omega_{g}$ do not depend on the magnetic quantum numbers $M'$ and $M$, respectively.
We note that the hfs splitting is small compared to the optical transition frequency. Therefore, we neglect the hfs splitting in our calculations. In other words, we assume that the energies $\hbar\omega_e$ and  $\hbar\omega_g$  do not depend on the quantum numbers $F'$ and $F$, respectively, that is, we use the approximations $\omega_e=\omega_{n'J'}$ and $\omega_g=\omega_{nJ}$.  
 
We introduce the notation $\mathbf{d}_{eg}=\langle e|\mathbf{D}|g\rangle$ for the dipole
matrix element of the transition $|e\rangle\leftrightarrow|g\rangle$, where $\mathbf{D}$ is the electric dipole operator.
In an arbitrary quantization coordinate system $\{x_Q,y_Q,z_Q\}$, 
the spherical tensor components $d^{(q)}_{eg}$ of the  dipole matrix element $\mathbf{d}_{eg}$, with the index $q=0,\pm1$,  are given by the expression \cite{Shore}
\begin{eqnarray}\label{a4}
	d^{(q)}_{eg}&=&(-1)^{I+J'-M'}\langle n'J' \| D\| nJ\rangle\sqrt{(2F+1)(2F'+1)}\nonumber\\
	&&\mbox{}\times
	\begin{Bmatrix}J'&F'&I\\F&J&1\end{Bmatrix}
	\begin{pmatrix}F&1&F'\\M&q&-M'\end{pmatrix}.
\end{eqnarray}
Here, the array in the curly braces is a 6$j$ symbol, the array in the parentheses is a 3$j$ symbol, $I$ is the nuclear spin, and $\langle n'J' \| D\| nJ\rangle$ is the reduced electric dipole matrix element in the $J$ basis.
Note that $d^{(q)}_{eg}$ is nonzero only if $M'-M=q=0,\pm1$.

The 3$j$ and 6$j$ symbols have the following orthogonal properties \cite{tensorbook}:
\begin{eqnarray}\label{a5}
&&\sum_{m_1m_2}
	\bigg(\begin{array}{ccc}
		j_1 &j_2  &j_3 \\
		m_1 &m_2 & m_3
	\end{array}\bigg)
\bigg(\begin{array}{ccc}
	j_1 &j_2  &j'_3 \\
	m_1 &m_2 & m'_3
\end{array}\bigg)\nonumber\\
&&\qquad =\frac{1}{2j_3+1}\delta_{j_3j'_3}\delta_{m_3m'_3}  \{j_1\;j_2\;j_3\}
\end{eqnarray}
and
\begin{eqnarray}\label{a6}
	&&\sum_{j_3}(2j_3+1)
	\bigg\{\begin{array}{ccc}
		j_1 &j_2  &j_3 \\
		j_4 &j_5 & j_6
	\end{array}\bigg\}
	\bigg\{\begin{array}{ccc}
	j_1 &j_2  &j_3 \\
	j_4 &j_5 & j'_6
\end{array}\bigg\}
\nonumber\\
	&&\qquad =\frac{\delta_{j_6j'_6}}{2j_6+1}\{j_1\;j_5\;j_6\}\{j_4\;j_2\;j_6\}.
\end{eqnarray}
Here, the triangular delta $\{j_1\; j_2\; j_3\}$ is equal to 1 when the triad $(j_1, j_2, j_3)$ satisfies the triangle conditions, and is zero otherwise.
With the help of Eqs.~(\ref{a5}) and (\ref{a6}), we find
\begin{eqnarray}\label{a7}
\frac{1}{2F+1}\sum_{MF'M'}\mathbf{d}_{eg}\cdot\mathbf{T}\cdot\mathbf{d}_{eg}^*&=&
\frac{|\langle n'J' \| D\| nJ\rangle|^2}{3(2J+1)}\mathrm{Tr}(\mathbf{T}),
\nonumber\\
\frac{1}{2F'+1}\sum_{M'FM}\mathbf{d}_{eg}\cdot\mathbf{T}\cdot\mathbf{d}_{eg}^*&=&
\frac{|\langle n'J' \| D\| nJ\rangle|^2}{3(2J'+1)}\mathrm{Tr}(\mathbf{T}),\quad		
\end{eqnarray}
where $\mathbf{T}$ is an arbitrary dyadic tensor. Equations (\ref{a7}) allow us to carry out analytically
the summation over the relevant sublevels.

Let $U_g$ and $U_e$ be the Casimir-Polder potentials of the lower state $|g\rangle=|nJFM\rangle$ and the upper state $|e\rangle=|n'J'F'M'\rangle$, respectively, induced by the transitions $|g\rangle\leftrightarrow|e\rangle$ between these two states. The total potential of the lower state $|nJFM\rangle$ or the upper state $|n'J'F'M'\rangle$,
caused by the transitions to the set of the sublevels of the upper term $|n'J'\rangle$ or the lower term $|nJ\rangle$, respectively, is given as $U_{nJFM}=\sum_{F'M'}U_g$ or $U_{n'J'F'M'}=\sum_{FM}U_e$. 
The average $\bar{U}_{\bar{g}}$ of the potentials $U_{nJFM}$ with respect to the magnetic quantum number $M$ and 
the average $\bar{U}_{\bar{e}}$ of
the potentials $U_{n'J'F'M'}$ with respect to the corresponding magnetic quantum number $M'$ are given as
\begin{eqnarray}\label{a8}
	\bar{U}_{\bar{g}}&=&\frac{1}{2F+1}\sum_{MF'M'} U_g,\nonumber\\
	\bar{U}_{\bar{e}}&=&\frac{1}{2F'+1}\sum_{M'FM} U_e.
\end{eqnarray}
Thus, $\bar{U}_{\bar{g}}$ and $\bar{U}_{\bar{e}}$ are the mean Casimir-Polder potentials of the magnetic sublevels of the hyperfine-structure levels $|\bar{g}\rangle=|nJF\rangle$ and $|\bar{e}\rangle=|n'J'F'\rangle$, having a flat statistical distribution over the magnetic sublevels. According to Ref.~\cite{Stourm2020}, the dependence of the potential of an individual magnetic sublevel $|nJFM\rangle$ or $|n'J'F'M'\rangle$ on $M$ or $M'$ is weak in the region of distances $r-a\geq 60$ nm. Hence, the differences between $\bar{U}_{\bar{g}}$ and $U_{nJFM}$ and between $\bar{U}_{\bar{e}}$ and $U_{n'J'F'M'}$ are  small when the atom-to-surface distance is not too short.

When we neglect the dependencies of the transition frequency $\omega_{eg}$ on $F$ and $F'$ and
use Eqs.~(\ref{a7}) to carry out analytically
the summation over the relevant sublevels, we find
\begin{eqnarray}\label{a9}
	\bar{U}_{\bar{g}}&=&\frac{\omega_{eg}}{3\pi\epsilon_0 c^2}\frac{|\langle n'J' \| D\| nJ\rangle|^2}{2J+1}
	\int\limits_0^{\infty}d u\frac{u^2}{\omega_{eg}^2+u^2}
	\nonumber\\&&\mbox{}
	\times\mathrm{Tr}[\mathbf{G}^{(\mathrm{sc})}(\mathbf{R},\mathbf{R};i u)],\nonumber\\	
	\bar{U}_{\bar{e}}&=&-\frac{\omega_{eg}}{3\pi\epsilon_0 c^2}\frac{|\langle n'J' \| D\| nJ\rangle|^2}{2J'+1}
	\int\limits_0^{\infty}d u\frac{u^2}{\omega_{eg}^2+u^2}
	\nonumber\\&&\mbox{}
	\times\mathrm{Tr}[\mathbf{G}^{(\mathrm{sc})}(\mathbf{R},\mathbf{R};i u)]\nonumber\\	
	&&\mbox{}-\frac{\omega_{eg}^2}{3\epsilon_0 c^2}\frac{|\langle n'J' \| D\| nJ\rangle|^2}{2J'+1}
	\mathrm{Re}\{\mathrm{Tr}[\mathbf{G}^{(\mathrm{sc})}(\mathbf{R},\mathbf{R};\omega_{eg})]\}.
	\nonumber\\
\end{eqnarray}
It is clear that expressions (\ref{a9}) for the mean potentials $\bar{U}_{\bar{g}}$ and $\bar{U}_{\bar{e}}$ do not contain the summation over the magnetic sublevels. Due to the symmetry of the atomic wave functions and the averaging procedure, these expressions contain the reduced electric dipole matrix element $\langle n'J' \| D\| nJ\rangle$ instead of the electric dipole vector $\mathbf{d}$. Furthermore, we see that the mean potentials $\bar{U}_{\bar{g}}$ and $\bar{U}_{\bar{e}}$ do not depend on the orientation of the quantization axis. Consequently, the diagonalization of the full Hamiltonian in the relevant Hilbert subspace is not necessary for the calculations of these mean potentials.

It is interesting to note that the mean Casimir-Polder potentials $\bar{U}_{\bar{g}}$ and $\bar{U}_{\bar{e}}$ do not depend on the quantum numbers $F$ and $F'$, respectively. Due to this property, $\bar{U}_{\bar{g}}$ or $\bar{U}_{\bar{e}}$ can be considered as the mean Casimir-Polder potential of the atom in the lower term $|\bar{\bar{g}}\rangle=|nJ\rangle$ or the upper term $|\bar{\bar{e}}\rangle=|n'J'\rangle$ in the framework of the two-term atom model. 

Comparison between Eqs.~(\ref{a1}) and (\ref{a9}) shows that the mean potential $\bar{U}_{\bar{g}}$ or $\bar{U}_{\bar{e}}$ can be formally considered as the potential of the ground or excited state of a two-level atom with an isotropically averaged dipole moment of magnitude
$|d|=|\langle n'J' \| D\| nJ\rangle|/\sqrt{2J+1}$ or $|d|=|\langle n'J' \| D\| nJ\rangle|/\sqrt{2J'+1}$, respectively. This effective dipole moment of the two-term atom depends on the reduced dipole matrix element of the transitions and the degeneracy of the initial-state term. It is clear that,  when $J\not=J'$, the effective dipoles for the potentials of the lower term $|\bar{\bar{g}}\rangle=|nJ\rangle$ and the upper term $|\bar{\bar{e}}\rangle=|n'J'\rangle$ have different magnitudes, unlike the case of two-level atoms [see Eqs.~(\ref{a1})]. This feature occurs because, when the numbers of the sublevels of the lower and upper terms are not the same, the mean dipole coupling strength for the sublevels of a term is different from that of the other term. 

We now consider the full set of fine-structure energy levels of the multilevel alkali-metal atom. 
It is clear that the Casimir-Polder potential $U$ of the atom in an arbitrary level $|a\rangle=|n_aJ_a\rangle$ is the sum of the contributions from the transitions between this level and other levels $|b\rangle=|n_bJ_b\rangle$, that is, 
\begin{eqnarray}\label{a11}
&& U = -\sum_{b}\frac{\omega_{ab}}{3\pi\epsilon_0 c^2}\frac{|\langle n_aJ_a \| D\| n_bJ_b\rangle|^2}{2J_a+1}
\int\limits_0^{\infty}d u\frac{u^2}{\omega_{ab}^2+u^2}
\nonumber\\&&\mbox{}
\times\mathrm{Tr}[\mathbf{G}^{(\mathrm{sc})}(\mathbf{R},\mathbf{R};i u)]\nonumber\\	
&&\mbox{}-\sum_{b}\Theta(\omega_{ab})\frac{\omega_{ab}^2}{3\epsilon_0 c^2}\frac{|\langle n_aJ_a \| D\| n_bJ_b\rangle|^2}{2J_a+1}
\nonumber\\&&\mbox{}
\times\mathrm{Re}\{\mathrm{Tr}[\mathbf{G}^{(\mathrm{sc})}(\mathbf{R},\mathbf{R};\omega_{ab})]\}. 
\end{eqnarray}
Here, $\omega_{ab}=\omega_a-\omega_b=-\omega_{ba}$ is the transition frequency and 
$\Theta(\omega_{ab})$ is the Heaviside step function of $\omega_{ab}$.
Expression (\ref{a11}) allows us to calculate the Casimir-Polder potential of the atom using the transition frequencies $\omega_{ab}$, the reduced dipole matrix elements $\langle n_aJ_a\| D\| n_bJ_b\rangle$, and the scattering Green tensor $\mathbf{G}^{(\mathrm{sc})}$.

In general, the potential $U$ can be decomposed as $U=U_{\mathrm{nres}}+U_{\mathrm{res}}$,  with the nonresonant part
\begin{eqnarray}\label{a13}
	U_{\mathrm{nres}}&=&
	-\sum_{b}\frac{\omega_{ab}}{3\pi\epsilon_0 c^2}\frac{|\langle n_aJ_a \| D\| n_bJ_b\rangle|^2}{2J_a+1}
	\int\limits_0^{\infty}d u\frac{u^2}{\omega_{ab}^2+u^2}
	\nonumber\\&&\mbox{}
	\times\mathrm{Tr}[\mathbf{G}^{(\mathrm{sc})}(\mathbf{R},\mathbf{R};i u)]	
\end{eqnarray}
and the resonant part
\begin{eqnarray}\label{a14}
	U_{\mathrm{res}}&=&
	-\sum_{b}\Theta(\omega_{ab})\frac{\omega_{ab}^2}{3\epsilon_0 c^2}\frac{|\langle n_aJ_a \| D\| n_bJ_b\rangle|^2}{2J_a+1}
\nonumber\\&&\mbox{}
\times\mathrm{Re}\{\mathrm{Tr}[\mathbf{G}^{(\mathrm{sc})}(\mathbf{R},\mathbf{R};\omega_{ab})]\}.
\end{eqnarray}
The presence of the Heaviside step function $\Theta(\omega_{ab})$ in expression (\ref{a14}) indicates that only downward transitions (with $\omega_{ab}>0$) can contribute to the resonant part $U_{\mathrm{res}}$ of the Casimir-Polder potential $U$.  

In the particular case where the state $|a\rangle$ of the multilevel atom is the ground fine-structure state $|n_gJ_g\rangle$, the resonant part $U_{\mathrm{res}}$ of the potential $U$ is vanishing. In this case,
we have $U=U_{\mathrm{nres}}$, that is,
\begin{eqnarray}\label{a15}
U&=&\sum_{e}\frac{\omega_{eg}}{3\pi\epsilon_0 c^2}\frac{|\langle n_eJ_e \| D\| n_gJ_g\rangle|^2}{2J_g+1}
\int\limits_0^{\infty}d u\frac{u^2}{\omega_{eg}^2+u^2}
\nonumber\\&&\mbox{}
\times\mathrm{Tr}[\mathbf{G}^{(\mathrm{sc})}(\mathbf{R},\mathbf{R};i u)],
\end{eqnarray}
where $|n_eJ_e\rangle$ is an arbitrary excited fine-structure state and $\omega_{eg}=\omega_{n_eJ_e}-\omega_{n_gJ_g}$ is the transition frequency. 
We can rewrite the above expression for the Casimir-Polder potential of a ground-state atom in the form \cite{Buhmann2012}
\begin{eqnarray}\label{a16}
	U&=&\frac{\hbar}{2\pi\epsilon_0 c^2}
	\int\limits_0^{\infty}d u\; u^2\alpha(iu)
	\mathrm{Tr}[\mathbf{G}^{(\mathrm{sc})}(\mathbf{R},\mathbf{R};i u)],\qquad
\end{eqnarray}
where
\begin{eqnarray}\label{a17}
	\alpha(\omega)
	&=&\frac{2}{3(2J_g+1)\hbar}\sum_{e}
	|\langle{n_e J_e\| D\|n_g J_g}\rangle|^2
	\frac{\omega_{eg}}{\omega_{eg}^2-\omega^2}\qquad
\end{eqnarray}
is the scalar polarizability of the ground-state alkali-metal atom \cite{Rosenbusch09,Fam12}. 

We emphasize that, in terms of the general scattering Green tensor $\mathbf{G}^{(\mathrm{sc})}$,
Eqs.~(\ref{a11})--(\ref{a16}) are valid for a multilevel alkali-metal atom in the presence of an arbitrary nonmagnetic dielectric or metallic body \cite{Buhmann2012}. In the particular case of a dielectric or metallic cylinder, an explicit expression for $\mathrm{Tr}[\mathbf{G}^{(\mathrm{sc})}(\mathbf{R},\mathbf{R};i u)]$ is given in Appendix \ref{sec:Green}. Equation (\ref{a16}) with $\mathrm{Tr}[\mathbf{G}^{(\mathrm{sc})}(\mathbf{R},\mathbf{R};i u)]$ given by Eq.~(\ref{m25}) is similar to but more rigorous than the results of Refs.~\cite{Boustimi2002,Boustimi2003,Frawley2012} for the nonretarded van der Waals potential of a ground-state atom near a dielectric or metallic cylinder.

\section{Numerical calculations}
\label{sec:num}

We perform numerical calculations for the Casimir-Polder potential of a multilevel alkali-metal atom near a vacuum-clad silica nanofiber  [see Fig.~\ref{fig1}(a)].
The nanofiber is an ultrathin fiber of radius $a$ and refractive index $n_1$ and is surrounded by an infinite background vacuum or air medium of refractive index $n_2=1$. The nanofiber diameters are a few hundreds of nanometers. 
We use Cartesian coordinates $\{x,y,z\}$, where $z$ is the coordinate along the fiber axis, and also cylindrical coordinates $\{r,\varphi,z\}$, where $r$ and $\varphi$ are the polar coordinates in the fiber transverse plane $xy$.  

To be concrete, we study a rubidium atom. 
The transition frequencies and reduced dipole matrix elements of atomic rubidium are taken from Ref.~\cite{rubidium}. Since these parameters are the same for rubidium-87 and rubidium-85, our numerical results are valid for both isotopes. 
We note that, according to Ref.~\cite{Stourm2020}, the contributions of the quadrupole transitions of the atom are not important for the potentials of the excited states with the principal quantum number $n\leq20$. Therefore, we limit our numerical calculations to the cases where the principal quantum number of the initial atomic state is $n\leq 10$.

The Green tensor of an infinitely long dielectric or metallic cylinder surrounded by a bulk dielectric medium is given in \cite{Tai1994,Chew1999,Li2000,Kornovan2016,Stourm2020}. The relevant expressions for the scattering part of the Green tensor are summarized in Appendix \ref{sec:Green}. 

The Green tensor depends on the permittivity of the nanofiber. The material of the nanofiber is fused silica. To calculate the permittivity of silica, we use the four-term Sellmeier formula, which is good for the spectral range from 200 nm to 7 $\mu$m \cite{Malitson,Ghosh}, and the Dawson-function model, which is good for the spectral range between 7 and 50 $\mu$m \cite{Meneses,Kitamura} (see also Appendix \ref{sec:Dawson}). 
Since the poles of the Green tensor, which characterize the resonances of the field, lie close to the real frequency axis, a special treatment of the integral path to avoid them is necessary in the calculations of $U_{\mathrm{res}}$ \cite{Fussell2004,Marocico2009,Ellingsen2010}.

\subsection{Potential for the ground state}

\begin{figure}[tbh]
	\begin{center}
		\includegraphics{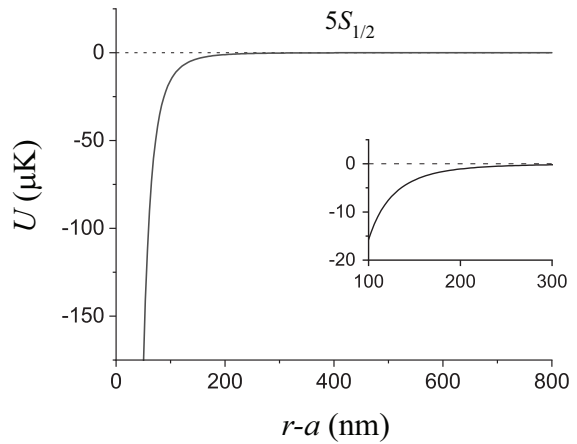}
	\end{center}
	\caption{ 
		Casimir-Polder potential $U$ of a rubidium atom in the ground state $5S_{1/2}$ versus the atom-to-surface distance $r-a$. The inset shows the details of the potential in a small range of the distance $r-a$. The radius of the nanofiber is $a=200$ nm.
	}
	\label{fig2}
\end{figure}

We calculate the Casimir-Polder potential $U$ of the rubidium atom in the ground state $5S_{1/2}$. In the calculations, we take into account the dominant dipole transitions from $5S_{1/2}$ to the neighboring levels $nP_{3/2}$ and $nP_{1/2}$ with $n=5$, 6, \dots 8. Our additional calculations, which are not shown here, confirm that the contributions of the levels with $n\geq9$ to the potential of the ground state are negligible.
Since the wavelengths of the dominant transitions from $5S_{1/2}$ are significantly smaller than 7 $\mu$m, we use the Sellmeier formula \cite{Malitson,Ghosh} to calculate the permittivity of silica. 

We plot in Fig.~\ref{fig2} the Casimir-Polder potential $U$ of the rubidium atom in the ground state $5S_{1/2}$ as a function of the distance $r-a$ from the atom to the fiber surface.
We observe from Fig.~\ref{fig2} that the Casimir-Polder potential $U$ of $5S_{1/2}$ is negative and monotonically reduces (increases) with decreasing (increasing) distance $r-a$. It is clear that the Casimir-Polder force $F=-\partial U/\partial r$ is negative, that is, the potential is attractive.
This behavior is in agreement with the results of the previous studies for the van der Waals potential of an alkali atom in the vicinity of a nanofiber \cite{twocolor,Frawley2012}. 

The monotonic falling-off behavior of the potential $U$ of the ground state is a consequence of the fact that this potential contains only the nonresonant terms [see Eq.~(\ref{a15})]. The distance dependencies of the nonresonant terms appear through the Green tensor $\mathbf{G}^{(\mathrm{sc})}(\mathbf{R},\mathbf{R};i u)$ at imaginary frequency. For the imaginary frequency $iu$, the tensor $\mathbf{G}^{(\mathrm{sc})}(\mathbf{R},\mathbf{R};i u)$ contains
the monotonically varying modified Bessel functions of the second kind $K_n(q_2r)$, 
where $q_2=\sqrt{u^2/c^2+\beta^2}$ is a real parameter with $\beta$ varying from $-\infty$ to $+\infty$  [see Eq.~(\ref{m25})].

\begin{figure}[tbh]
	\begin{center}
		\includegraphics{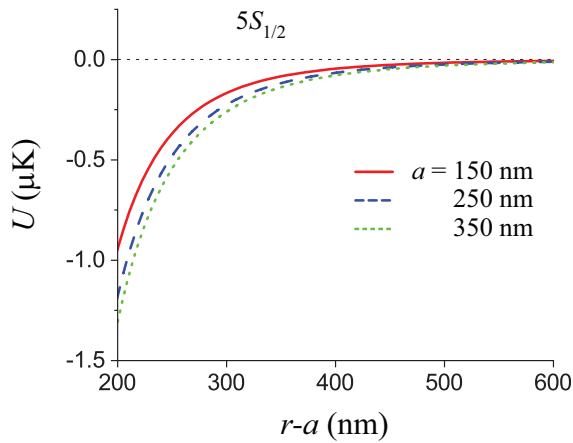}
	\end{center}
	\caption{ 
		Comparison between the Casimir-Polder potentials $U$ of a rubidium atom in the ground state $5S_{1/2}$ for the fiber radii  $a=150$, 250, and 350  nm.
	}
	\label{fig3}
\end{figure}

In Fig.~\ref{fig3}, we compare the Casimir-Polder potentials $U$ of the ground state $5S_{1/2}$ for different values of the fiber radius $a$. The figure shows that, in the range of the parameters used for calculations, a moderate variation of the fiber radius does not affect much the monotonic falling-off behavior of the potential $U$ of the ground state. It is clear from the figure that an increase of the fiber radius leads to an increase of the absolute value of the potential.

\subsection{Potentials for the excited states of the $D_2$ and $D_1$ lines}

The first two excited states of the rubidium atom are the states $5P_{3/2}$ and $5P_{1/2}$.
The only downward electric dipole transitions from these excited states are the transitions to the ground state $5S_{1/2}$. The transitions $5P_{3/2}\leftrightarrow5S_{1/2}$ and $5P_{1/2}\leftrightarrow5S_{1/2}$  are called the $D_2$ and $D_1$ lines, respectively. They play an important role in laser cooling and atom--light interaction experiments \cite{coolingbook}. Below, we present the numerical results for the Casimir-Polder potentials of the atom in these  excited states.

In the calculations for the potentials of $5P_{3/2}$ and $5P_{1/2}$, we take into account the dominant dipole transitions to the states $nS_{1/2}$ with $n=5$, 6, \dots 8 and the states $nD_{3/2}$  with $n=4$, 5, \dots 8. For the potential of the state $5P_{3/2}$, we also include the dominant dipole transitions to the states $nD_{5/2}$ with $n=4$, 5, \dots 8. The contributions of the levels with $n\geq9$ to the potentials of $5P_{3/2}$ and  $5P_{1/2}$ are negligible. Since the wavelengths of the dominant transitions from $5P_{3/2}$ and $5P_{1/2}$  are significantly smaller than 7 $\mu$m, the permittivity of silica  is calculated using the Sellmeier formula \cite{Malitson,Ghosh}.

\begin{figure}[tbh]
	\begin{center}
		\includegraphics{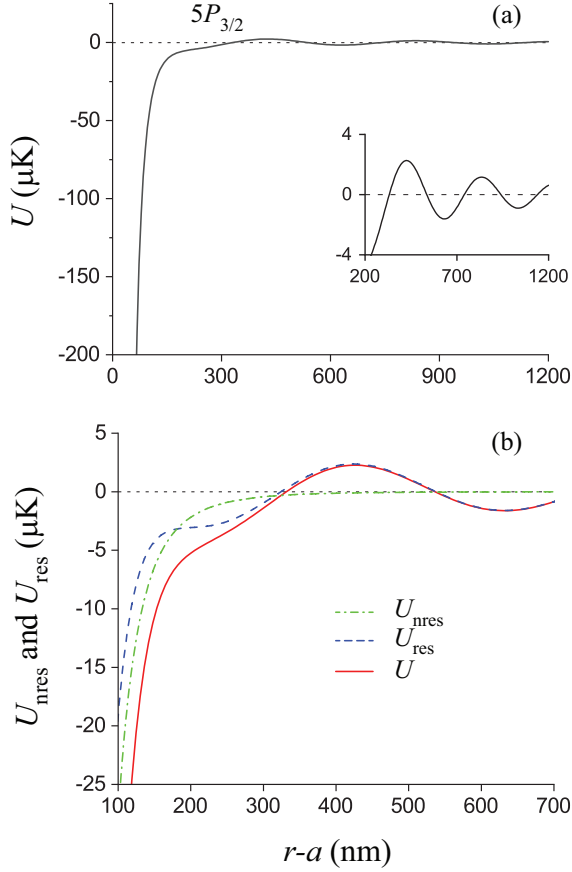}
	\end{center}
	\caption{ 
		(a) Casimir-Polder potential $U$ of a rubidium atom in the excited state $5P_{3/2}$ of the $D_2$ line versus the atom-to-surface distance $r-a$. 		
		(b) Nonresonant part $U_{\mathrm{nres}}$ (dash-dotted green line) and resonant part $U_{\mathrm{res}}$  (dashed blue line) of the potential $U$ (solid red line). 
		In parts (a) and (b), the radius of the nanofiber is $a=200$ nm.
	}
	\label{fig4}
\end{figure}

We plot in Fig.~\ref{fig4}(a) the Casimir-Polder potential $U$ for the rubidium atom in the excited state $5P_{3/2}$ of the $D_2$ line as a function of the distance $r-a$ from the atom to the fiber surface.  We display in Fig.~\ref{fig4}(b) the nonresonant part $U_{\mathrm{nres}}$ and the resonant part $U_{\mathrm{res}}$ of the potential $U$. 
 
We observe from Fig.~\ref{fig4}(a) that, in the region of short distances, the Casimir-Polder potential $U$ of the rubidium atom in the excited state $5P_{3/2}$ is negative, monotonic, and attractive. However, in the region of moderate and large distances, the potential $U$ becomes positive in some spatial intervals and oscillates around the zero value. The wavelength of the spatial oscillations of $U$ is about half of the wavelength of 780 nm of the $D_2$ line of atomic rubidium \cite{coolingbook}. Due to the spatial oscillations, the Casimir-Polder potential $U$ can become repulsive, that is, the corresponding force $F=-\partial U/\partial r$ can become positive, depending on the distance. It is clear that this effect occurs in the retardation regime. The maximal amplitude of oscillations of the potential is about $2.3$ $\mu$K. The maximal depth of the wells created by the spatial oscillations of the potential is about $2.8$ $\mu$K. This depth is about 15 times larger than the recoil energy $E_{\mathrm{r}}/k_\mathrm{B} \cong  181$ nK for the $D_2$ line of atomic rubidium \cite{coolingbook}. 

Note that similar repulsive Casimir forces between surfaces in a liquid  were obtained for suitable choices of surfaces \cite{repulsive}. An oscillatory surface-induced potential was reported  in Ref.~\cite{chiralforcefull} for 
the excited state of a two-level atom near a nanofiber.  

Comparison between Figs.~\ref{fig2} and \ref{fig4}(a) for the potentials of the ground state $5S_{1/2}$ and the excited state $5P_{3/2}$ shows that, in the region of short distances, both potentials are attractive but the magnitude of the potential for $5P_{3/2}$ is larger than that for $5S_{1/2}$.

The oscillations in the spatial dependence of the Casimir-Polder potential $U$ of the excited state
$5P_{3/2}$ originate from the resonant part $U_{\mathrm{res}}$ [see the dashed blue curve in Fig.~\ref{fig4}(b)] and appear through the spatial dependence of the Green tensor $\mathbf{G}^{(\mathrm{sc})}(\mathbf{R},\mathbf{R};\omega_{ab})$ at the resonant transition frequency $\omega_{ab}$ [see Eq.~(\ref{a14})]. This tensor contains oscillating Hankel functions of the first kind $H_n^{(1)}(\eta_2 r)$, where $\eta_2=\sqrt{\omega^2/c^2-\beta^2}$ is a real parameter for the radiation modes with the longitudinal propagation constant $|\beta|\leq \omega/c$ [see Eq.~(\ref{m24})]. The spatial oscillations of the potential $U$ can be ascribed to the constructive/destructive interference between the quantum light waves emitted from the atom and reflected from the fiber surface. The wavelength of the spatial oscillations of the potential is close to  half the wavelengths of the dominant downward resonant transitions of the atom. 

We observe from Fig.~\ref{fig4}(b) that, for increasing distance $r-a$, the magnitude of the nonresonant component $U_{\mathrm{nres}}$ of the Casimir-Polder potential $U$ reduces quickly and monotonically approaches zero. Meanwhile, the magnitude of the resonant component $U_{\mathrm{res}}$ first reduces quickly but then starts oscillating around the zero value with a decaying amplitude. In the region $r-a > 180$ nm in the case of the figure, $U_{\mathrm{res}}$ is dominant over $U_{\mathrm{nres}}$ and, hence, the total potential $U$ is determined mainly by $U_{\mathrm{res}}$. However, in the region $r-a < 180$ nm in the case of the figure, the magnitude of the resonant component of the  potential is smaller than that of the nonresonant component. This feature is a consequence of the multilevel structure of the atom, where all  higher levels can contribute to the nonresonant component of the Casimir-Polder  potential  \cite{Buhmann2004}.

\begin{figure}[tbh]
	\begin{center}
		\includegraphics{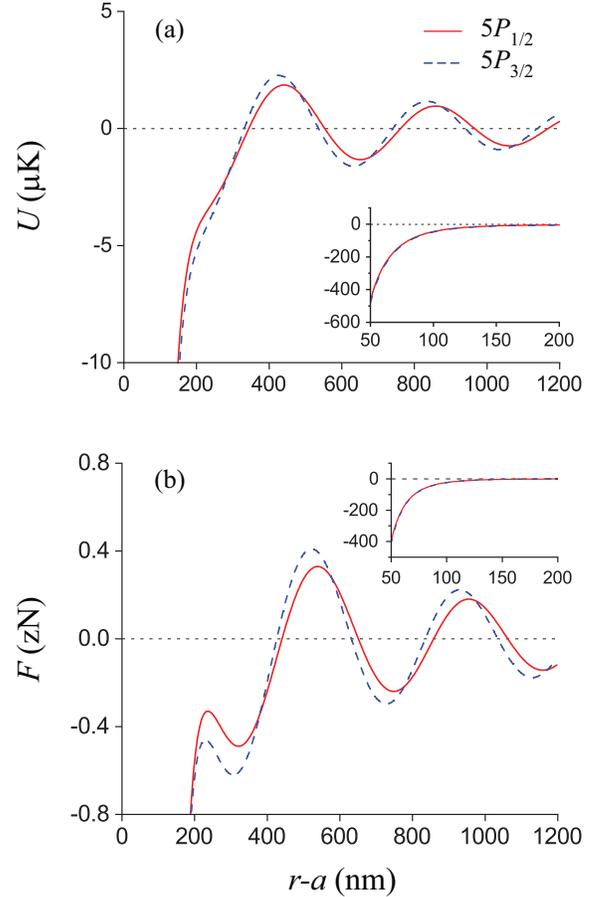}
	\end{center}
	\caption{ 
		Casimir-Polder potentials $U$ (a) and the corresponding forces $F=-\partial U/\partial r$ (b) for a rubidium atom in the excited state $5P_{1/2}$ of the $D_1$ line (solid red line) and the excited state $5P_{3/2}$ of the $D_2$ line (dashed blue line) versus the atom-to-surface distance $r-a$. The radius of the nanofiber is $a=200$ nm.
	}
	\label{fig5}
\end{figure}

We plot in Fig.~\ref{fig5}(a) the potential of the excited state $5P_{1/2}$ of the $D_1$ line. For comparison, we also plot in this figure the potential of the excited state $5P_{3/2}$  of the $D_2$ line.
We observe from the figure that the potentials for $5P_{1/2}$ and $5P_{3/2}$  are very similar to each other. Both are attractive for the short distances and are oscillatory for the distances in the range $r-a\geq 400$ nm. 
One of the reason for the similarity between the two potentials is that
the wavelength of the $D_1$ line (795 nm), which is responsible for the spatial oscillations of the potential of $5P_{1/2}$, is just slightly different from the wavelength of the $D_2$ line (780 nm), which is responsible for the spatial oscillations of the potential of $5P_{3/2}$.
Another reason is that the normalized reduced dipole matrix elements $|\langle a \| D\| b\rangle|/\sqrt{2J_a+1}$
for the downward transitions from the excited states $|a\rangle=5P_{1/2}$ and $5P_{3/2}$ to the ground state $|b\rangle=5S_{1/2}$ have the same value ($\simeq 2.98$ a.u.) \cite{rubidium}. 
The reduced dipole matrix elements and wavelengths of the $D_1$ and $D_2$ lines determine the resonant parts $U_{\mathrm{res}}$  [see Eq.~(\ref{a14})] of the potentials $U$ for $5P_{1/2}$ and $5P_{3/2}$.

The radial dependence of the Casimir-Polder potential $U$ leads to a radial Casimir-Polder force $F=-\partial U/\partial r$ acting on the center-of-mass motion of the atom. We plot in Fig.~\ref{fig5}(b) the radial dependencies of the forces $F$ on the atom in the excited states $5P_{1/2}$ and $5P_{3/2}$.
The figure shows that the forces can take positive values, depending on the region of space, and oscillate around the zero value with a decaying amplitude. The maximal positive value of the forces is on the order of $0.4$ zN. This value is about 40 times smaller than the maximal value $F_{\mathrm{sp}}^{(\mathrm{max})}=\hbar k\gamma/2\cong 15$ or 16 zN of the force resulting from absorption  followed by spontaneous emission for the $D_1$ or $D_2$ line, respectively, of atomic rubidium in free space \cite{coolingbook}. Here, $k$ is the wave number of the resonant light and $\gamma$ is the linewidth of the atom.

\begin{figure}[tbh]
	\begin{center}
		\includegraphics{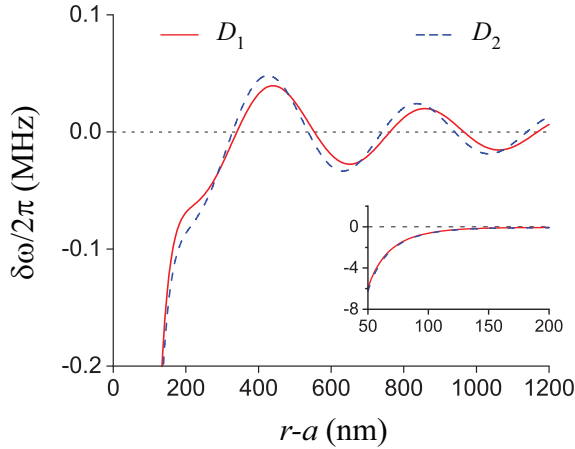}
	\end{center}
	\caption{ 
		Nanofiber-induced shifts $\delta\omega$ of the transition frequencies of the $D_1$  (solid red curve) 
		and  $D_2$  (dashed blue curve) lines of atomic rubidium versus the atom-to-surface distance $r-a$. 
		The radius of the nanofiber is $a=200$ nm.
	}
	\label{fig6}
\end{figure}

The difference between the Casimir-Polder potentials $U_a=U(|a\rangle)$ and $U_b=U(|b\rangle)$ of the levels $|a\rangle$ and $|b\rangle$ determines the shift $\delta\omega=(U_a-U_b)/\hbar$ of the atomic transition frequency $\omega_{ab}=\omega_a-\omega_b$.
We plot in Fig.~\ref{fig6} the nanofiber-induced shifts $\delta\omega$ of the transition frequencies of the $D_1$  and  $D_2$ lines of atomic rubidium.
The figure shows that the frequency shifts of these lines are almost the same. The shifts are negative in the region of short distances and oscillate around the zero value with a decaying amplitude for the distances in the range from 400 to 1200 nm. The shifts can become positive depending on the distance. The oscillatory behavior of the frequency shifts $\delta\omega$ shown in Fig.~\ref{fig6} occurs as a consequence of the oscillatory behavior of the Casimir-Polder potentials of the excited states $5P_{1/2}$ and $5P_{3/2}$, and is a signature of the retardation and interference effects.

\begin{figure}[tbh]
	\begin{center}
		\includegraphics{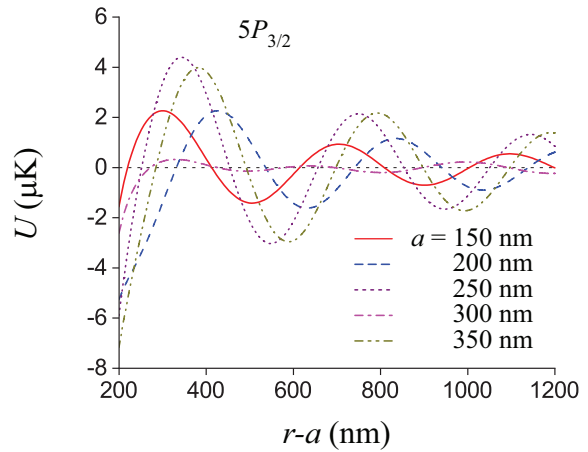}
	\end{center}
	\caption{ 
		Comparison between the Casimir-Polder potentials $U$ of a rubidium atom in the excited state $5P_{3/2}$ for the fiber radii  $a=150$, 200, 250, 300, and 350  nm.
	}
	\label{fig7}
\end{figure}

We compare in Fig.~\ref{fig7} the Casimir-Polder potentials $U$ of the excited state $5P_{3/2}$ for different values of the fiber radius $a$. The figure shows that 
the oscillatory behavior of $U$ is sensitive to the variation of $a$. 
Indeed, a minor or moderate variation of $a$ may cause a significant variation of oscillations of $U$. In particular, an increase of $a$ may lead to an increase or a decrease of the amplitude of oscillations of $U$, depending on the range of $a$. Among the curves of Fig.~\ref{fig7}, the curves for $a=250$ and 350 nm have the largest oscillation amplitudes (the peak values are on the order of 4 $\mu$K).

\subsection{Potentials for a few low-lying excited states $nS_{1/2}$}

\begin{figure}[tbh]
	\begin{center}
		\includegraphics[width=0.45\textwidth]{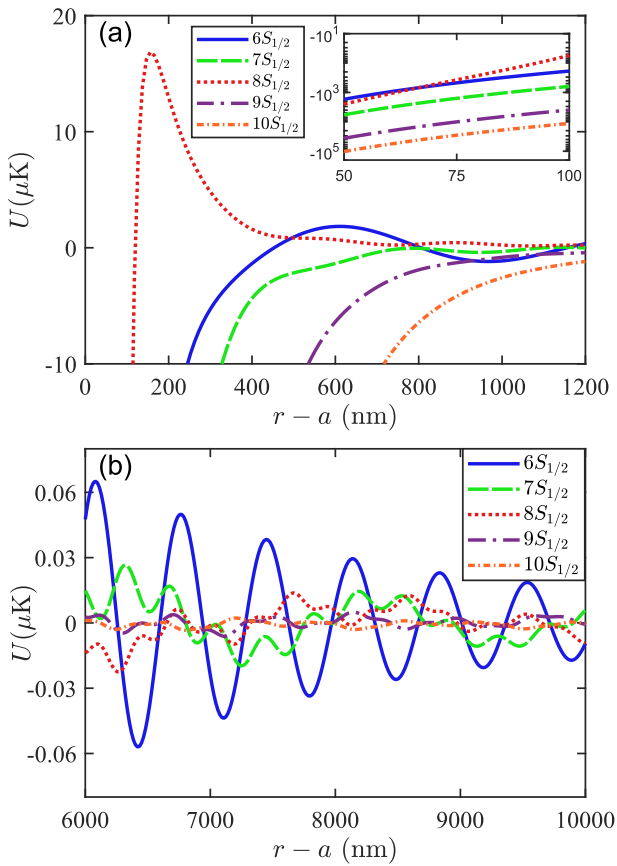}
	\end{center}
	\caption{(a) Casimir-Polder potentials $U$ for the states $nS_{1/2}$ with $n=6$, 7, \dots 10 versus the atom-to-surface distance $r-a$ in the range from 100 to 1200 nm. The logarithm scale of the vertical axis of the inset shows the rapid decrease of the magnitudes of the potentials in the range of short distances from 50 to 100 nm. 	The radius of the nanofiber is $a=200$ nm. (b) Same as (a) but for large distances in the range from 6 to 10 $\mu$m.}
	\label{fig8}
\end{figure}

We plot in Fig.~\ref{fig8} the Casimir-Polder potentials of several low-lying excited states $nS_{1/2}$ with $n=6$, 7, \dots 10. 
In the calculations of these potentials, we take into account the dominant dipole transitions to the neighboring levels $nP_{3/2}$ and $nP_{1/2}$ with $n=5$, 6, \dots 20. Our additional calculations, which are not shown here, confirm that the contributions of the levels $nP_{3/2}$ and $nP_{1/2}$ with $n\geq15$ to the potentials are negligible.
Since the wavelengths of the dominant transitions from the states $nS_{1/2}$ with $n=6$, 7, \dots 10  are larger than  1 $\mu$m, the permittivity of silica is calculated by using the Dawson-function model \cite{Meneses,Kitamura}.

The logarithmic scale of the vertical axis of the inset of Fig.~\ref{fig8}(a) shows that, in the range of short distances (from 50 to 100 nm), the magnitudes of the potentials for all of the aforementioned states rapidly decrease with increasing distance. We also observe from the inset that,  in the region of short distances, the potential for the state $8S_{1/2}$ shows a slightly different behavior than the rest: the curve for the state $8S_{1/2}$ (the dotted red line) is higher than the curve for the state $7S_{1/2}$ (the dashed green line) and crosses the curve for the state $6S_{1/2}$ (the solid blue line).
Except the state $8S_{1/2}$, the states $nS_{1/2}$ with $n\leq 10$ demonstrate the general tendency: the higher state $nS_{1/2}$ has the deeper potential in the region of short distances.
 
We observe from Fig.~\ref{fig8}(a) that, in the region of moderate distances from 100 to 1200 nm, the curves for the potentials behave in a very different manner. Namely, the potentials for some of these states demonstrate a nonmonotonic  behavior. Specifically, the potential for the state $8S_{1/2}$ undergoes a change in sign, achieves a  positive peak value $U_{\mathrm{max}}\cong17$ $\mu$K at a distance $r-a \cong 150$ nm, and becomes rather strongly repulsive in the region of distances from 150 to 400 nm. Furthermore, the potentials for the states $nS_{1/2}$ with $n=6$, 7, and 8 demonstrate a pronounced oscillatory behavior at the distances $r-a \ge 400$ nm [see Fig.~\ref{fig8}(a)], and these oscillations persist even when the atom is several micrometers away from the fiber surface, with an amplitude on the order of  10 nK [see Fig.~\ref{fig8}(b)]. The potentials for the states $9S_{1/2}$
and $10S_{1/2}$ are monotonic for the distances $r-a \le 1200$ nm [see Fig.~\ref{fig8}(a)] but have small oscillations with a small amplitude on the order of 1 nK for the distances from 6 to 10 $\mu$m [see Fig.~\ref{fig8}(b)].   

One can notice that, for the state $6 S_{1/2}$, the oscillatory behavior of the potential is caused by the contributions from the downward transitions $6S_{1/2}\to 5P_{1/2}$ and $6S_{1/2}\to 5P_{3/2}$, which have similar transition frequencies (with the wavelengths of 1324 and 1367 nm). However, for the higher states $7S_{1/2}$, $8S_{1/2}$,  $9S_{1/2}$, and  $10S_{1/2}$, the oscillatory behavior of the potential is a result of the beating between the contributions from several downward transitions with significantly different transition wavelengths.
When the atom is excited highly enough  (the principal quantum number is $n\geq9$ in the case of rubidium), due to the beating between a large number of downward transitions and the increase to the wavelengths of the dominant downward transitions, the oscillatory behavior of the Casimir-Polder potential becomes less prominent and may even practically disappear. We note that the potentials of highly excited Rydberg states, calculated in Ref.~\cite{Stourm2020}, are not oscillatory.  

The reason why the potential $U$ for the state $8S_{1/2}$ is so different from the potentials for the other states [see Fig.~\ref{fig8}(a) and the inset] is that $8S_{1/2}$ is the only state for which the resonant part $U_{\text{res}}$ is positive for $r-a < 1200$ nm. This leads to a slightly different behavior of the potential at the positions that are very close to the surface (see the inset), and to the presence of a well distinct ``bump'' at the distance $r-a \cong 150$ nm [see Fig.~\ref{fig8}(a)]. A deeper inspection of the properties of $8S_{1/2}$ reveals that the strongest downward transitions from this state are the transitions to the lower states $7P_{1/2}$ and $7P_{3/2}$, with the wavelengths of $\lambda_1=8.249$ $\mu$m and $\lambda_2=8.495$ $\mu$m, respectively \cite{rubidium}. For these wavelengths,  the real part of the permittivity $\epsilon$ of silica is negative and small and the associated imaginary part is significant, that is, silica is a lossy metal. In addition, we have $|\epsilon(\lambda_1)|\cong 0.69<1$ and $|\epsilon(\lambda_2)|\cong 1.07>1$. Meanwhile, it follows from Eq.~(153) of Ref.~\cite{Buhmann2004} that, for an atom situated above a semi-infinite half space containing a homogeneous medium, the resonant part of the Casimir-Polder potential is positive for short distances under the condition $|\epsilon|<1$. This condition is satisfied for $\lambda_1$ but is slightly violated for $\lambda_2$. Nevertheless, due to the curvature and the finite size of the cylindrical surface, the resonant component of the nanofiber-induced potential is positive for both wavelengths $\lambda_1$ and $\lambda_2$ in the region of short distances.

\begin{figure}[tbh]
	\begin{center}
		\includegraphics[width=0.45\textwidth]{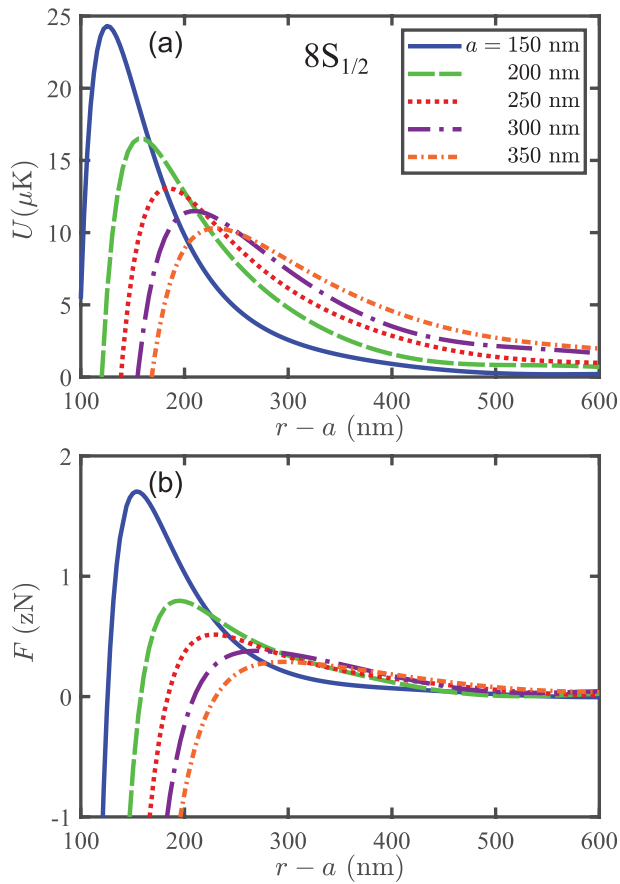}
	\end{center}
	\caption{ 
		Casimir-Polder potential $U$ (a) and corresponding force $F$ (b) for a rubidium atom in the excited state $8S_{1/2}$ as functions of the atom-to-surface distance $r-a$ for different fiber radii  $a=150$, 200, 250, 300, and 350 nm.
	}
	\label{fig9}
\end{figure}

In order to see the effect of the fiber radius $a$ on the behavior of the bump of the potential in the region of short distances, we plot in Fig.~\ref{fig9} the potential and the corresponding force for the excited state $8S_{1/2}$ as functions of the atom-to-surface distance $r-a$ for different values of $a$. The figure shows that the height  and  position of the bump are sensitive to the variation of $a$. For increasing $a$ in the range from 150 to 350 nm, the height of the bump decreases and its position moves away from the fiber surface. It is interesting to note that, for the parameters in the range considered, the smaller fiber radius leads to the stronger repulsion in the region of short distances. The reason is that the repulsion of the potential originates from the resonant part, which may increase faster than the nonresonant part decreases with reducing fiber radius.

\section{Summary}
\label{sec:summary}

In this paper, we have studied the Casimir-Polder potential of a multilevel alkali-metal atom near an optical nanofiber. 
We have calculated the mean potential of the atom in a fine-structure level. We have used the sum rules for $3j$ and $6j$ symbols to carry out analytically the summations over the relevant sublevels.

We have performed numerical calculations for the Casimir-Polder potentials of the ground state and a few low-lying excited states of rubidium. We have observed that the Casimir-Polder potential of the ground state $5S_{1/2}$ is negative  and its absolute value monotonically reduces to zero with increasing atom-to-surface distance. We have shown that the Casimir-Polder potential of a low-lying excited state ($5P_{1/2}$, $5P_{3/2}$, or $nS_{1/2}$ with $n=6$, 7, or 8) may take positive values, oscillate around the zero value with a decaying amplitude, and become repulsive in some regions of atom-to-surface distances. The spatial oscillations of the potential of a low-lying excited state occur as a consequence of the retardation and the interference between the emitted and reflected waves.
The characteristic wavelengths of the spatial oscillations of the potential are determined by the wavelengths of the dominant downward resonant transitions of the atom. 
The maximal amplitude of oscillations of the potential is on the order of a few microkelvins.
 
We have observed that, for a nanofiber with a radius of 200 nm, the potential for the state $8S_{1/2}$ of a rubidium atom achieves a  positive peak value of about 17 $\mu$K at a distance of about 150 nm from the fiber surface, and becomes rather strongly repulsive for distances from 150 to 400 nm. When the atom is initially in an  excited state with a principal quantum number $n\geq9$ in the case of rubidium, the oscillatory behavior of the Casimir-Polder potential is no longer prominent and may even disappear. This arises from the beating between the large number
of downward transitions and the increase to the wavelengths of the most significant downward transitions. 
 
We have also calculated the nanofiber-induced  shifts of the transition frequencies  of the $D_2$ and $D_1$ lines of atomic rubidium. We have found that the shifts are negative in the region of short distances, become positive, and oscillate around the zero value with a decaying amplitude in the region of large distances.  

Our results give insight into the Casimir-Polder potential of a multilevel alkali-metal atom near an optical nanofiber.
The repulsive character and spatially oscillatory behavior of the potential 
of a low-lying excited state may cause significant effects on the center-of-mass motion of cold atoms with a temperature comparable to or less than the oscillation amplitude, that is, on the order of or less than 1 $\mu$K.
Such temperatures can be achieved for atoms using the Raman cooling technique \cite{coolingbook}.

\begin{acknowledgments}
This work was supported by the Okinawa Institute of Science and Technology (OIST) Graduate University and by the Japan Society for the Promotion of Science (JSPS) Grant-in-Aid for Scientific Research (C) under Grants 19K05316 and 20K03795.
\end{acknowledgments}


\appendix

\begin{widetext}
	
	\section{Green tensor for a dielectric or metallic cylinder}
	\label{sec:Green}
	
	We consider an infinitely long, nonmagnetic, dielectric or metallic cylinder of radius $a$ and dielectric permittivity $\epsilon_1$ in a nonmagnetic dielectric medium of dielectric permittivity $\epsilon_2$. We use the Cartesian coordinates $\{x,y,z\}$, where $z$ is the coordinate along the cylinder axis.
	We also use the cylindrical coordinates $\{r,\varphi,z\}$, where $r$ and $\varphi$ are the polar coordinates in the transverse plane $xy$.
	When the field point $\mathbf{R}$ and the source point $\mathbf{R}'$ are outside the cylinder ($r>a$ and $r'>a$), we can decompose the Green tensor (the dyadic Green function) $\mathbf{G}(\mathbf{R},\mathbf{R}',\omega)$ as \cite{Buhmann2012,Tai1994,Chew1999,nano-optics}
	\begin{equation}\label{m5}
		\mathbf{G}(\mathbf{R},\mathbf{R}',\omega)=
		\mathbf{G}^{(0)}(\mathbf{R},\mathbf{R}',\omega)+\mathbf{G}^{(\mathrm{sc})}(\mathbf{R},\mathbf{R}',\omega).
	\end{equation}
	Here, $\mathbf{G}^{(0)}(\mathbf{R},\mathbf{R}',\omega)$ is the homogeneous-medium Green tensor and $\mathbf{G}^{(\mathrm{sc})}(\mathbf{R},\mathbf{R}',\omega)
	$ is the scattering Green tensor.
	
	The homogeneous-medium Green tensor $\mathbf{G}^{(0)}$ can be represented in the form \cite{Buhmann2012,nano-optics}
	\begin{equation}\label{m14}
		\mathbf{G}^{(0)}(\mathbf{R}_i,\mathbf{R}_j,\omega)=-\frac{1}{3k_2^2}\delta(\mathbf{R}_i-\mathbf{R}_j)\mathbf{I}
		+\frac{\exp(ik_2R_{ij})}{4\pi R_{ij}}\bigg[\bigg(1+\frac{ik_2R_{ij}-1}{k_2^2R_{ij}^2}\bigg)\mathbf{I}
		+\frac{3-3ik_2R_{ij}-k_2^2R_{ij}^2}{k_2^2R_{ij}^2}\hat{\mathbf{R}}_{ij}\hat{\mathbf{R}}_{ij}\bigg],
	\end{equation}
	where $k_2=\sqrt{\epsilon_2}\,\omega/c$. In Eq.~(\ref{m14}), the notations  $R_{ij}=|\mathbf{R}_{ij}|$ and $\hat{\mathbf{R}}_{ij}=\mathbf{R}_{ij}/R_{ij}$
	with $\mathbf{R}_{ij}=\mathbf{R}_i-\mathbf{R}_j$ have been used.
	
	Meanwhile, the scattering Green tensor $\mathbf{G}^{(\mathrm{sc})}$ can be given as \cite{Tai1994,Chew1999,Li2000,Kornovan2016,Stourm2020}
	\begin{eqnarray}\label{m15}
		\mathbf{G}^{(\mathrm{sc})}(\mathbf{R},\mathbf{R}',\omega)
		&=&\frac{i}{8\pi}\sum_{n=-\infty}^\infty\int_{-\infty}^{\infty}\frac{d\beta}{\eta_2^2}
		\{[A_R\mathbf{M}_{n\eta_2}^{(1)}(\beta,\mathbf{R})
		+B_R\mathbf{N}_{n\eta_2}^{(1)}(\beta,\mathbf{R})]
		\overline{\mathbf{M}}_{n\eta_2}^{(1)}(\beta,\mathbf{R}')
		\nonumber\\&&\mbox{}
		+[C_R\mathbf{N}_{n\eta_2}^{(1)}(\beta,\mathbf{R})
		+D_R\mathbf{M}_{n\eta_2}^{(1)}(\beta,\mathbf{R})]
		\overline{\mathbf{N}}_{n\eta_2}^{(1)}(\beta,\mathbf{R}')\}.
	\end{eqnarray}
	Here we have introduced the notations $k_j=\sqrt{\epsilon_j}\,\omega/c$ and 	 
	$\eta_j=\sqrt{k_j^2-\beta^2}=\sqrt{\epsilon_j\omega^2/c^2-\beta^2}$ for $j=1,2$. We have also used the vector wave functions
	\begin{eqnarray}\label{m16}
		\mathbf{M}_{n\eta}^{(1)}(\beta,\mathbf{R})&=&
		\Big[\frac{in}{r}H_n^{(1)}(\eta r)\hat{\mathbf{r}}
		-\eta H_n^{(1)\prime}(\eta r)\hat{\boldsymbol{\varphi}}\Big]e^{in\varphi+i\beta z}
		,\nonumber\\
		\mathbf{N}_{n\eta}^{(1)}(\beta,\mathbf{R})&=&
		\frac{1}{\sqrt{\eta^2+\beta^2}}\Big[
		i\beta\eta H_n^{(1)\prime}(\eta r)\hat{\mathbf{r}}
		-\frac{n\beta}{r}H_n^{(1)}(\eta r)\hat{\boldsymbol{\varphi}}
		+\eta^2H_n^{(1)}(\eta r)\hat{\mathbf{z}}
		\Big]e^{in\varphi+i\beta z},
	\end{eqnarray}
	\begin{eqnarray}\label{m17}
		\overline{\mathbf{M}}_{n\eta}^{(1)}(\beta,\mathbf{R})&=&
		\Big[-\frac{in}{r}H_n^{(1)}(\eta r)\hat{\mathbf{r}}
		-\eta H_n^{(1)\prime}(\eta r)\hat{\boldsymbol{\varphi}}\Big]e^{-in\varphi-i\beta z}
		,\nonumber\\
		\overline{\mathbf{N}}_{n\eta}^{(1)}(\beta,\mathbf{R})&=&
		\frac{1}{\sqrt{\eta^2+\beta^2}}\Big[
		-i\beta\eta H_n^{(1)\prime}(\eta r)\hat{\mathbf{r}}
		-\frac{n\beta}{r}H_n^{(1)}(\eta r)\hat{\boldsymbol{\varphi}}
		+\eta^2H_n^{(1)}(\eta r)\hat{\mathbf{z}}
		\Big]e^{-in\varphi-i\beta z},
	\end{eqnarray}
where $H_n^{(1)}$ is the Hankel function of the first kind.
	The coefficients $A_R$, $B_R$, $C_R$, and $D_R$ are given as \cite{Tai1994,Chew1999,Li2000,Kornovan2016,Stourm2020}
\begin{equation}\label{m22}
	\begin{split}
		A_R&=\frac{1}{W_R}\frac{J_n(\eta_2 a)}{H_n^{(1)}(\eta_2 a)}\bigg[\frac{n^2\beta^2}{a^2} \bigg(\frac{1}{\eta_2^2}-\frac{1}{\eta_1^2} \bigg)^2
		-\bigg(\frac{J'_n(\eta_1 a)}{\eta_1 J_n(\eta_1 a)}-\frac{J'_n(\eta_2 a)}{\eta_2 J_n(\eta_2 a)} \bigg)
		\bigg(\frac{k_1^2J'_n(\eta_1 a)}{\eta_1 J_n(\eta_1 a)}-\frac{k_2^2H^{(1)\prime}_n(\eta_2 a)}{\eta_2 H^{(1)}_n(\eta_2 a)} \bigg)\bigg],\\
		C_R&=\frac{1}{W_R}\frac{J_n(\eta_2 a)}{H_n^{(1)}(\eta_2 a)}\bigg[\frac{n^2\beta^2}{a^2} \bigg(\frac{1}{\eta_2^2}-\frac{1}{\eta_1^2} \bigg)^2
		-\bigg(\frac{J'_n(\eta_1 a)}{\eta_1 J_n(\eta_1 a)}-\frac{H^{(1)\prime}_n(\eta_2 a)}{\eta_2 H^{(1)}_n(\eta_2 a)} \bigg)
		\bigg(\frac{k_1^2J'_n(\eta_1 a)}{\eta_1 J_n(\eta_1 a)}-\frac{k_2^2J'_n(\eta_2 a)}{\eta_2 J_n(\eta_2 a)} \bigg)\bigg],\\
		B_R&=D_R=\frac{1}{W_R}\frac{J_n(\eta_2 a)}{H_n^{(1)}(\eta_2 a)}\frac{k_2}{\eta_2}\frac{n\beta}{a} \bigg(\frac{1}{\eta_2^2}-\frac{1}{\eta_1^2} \bigg)\bigg(\frac{J'_n(\eta_2 a)}{J_n(\eta_2 a)}-\frac{H^{(1)\prime}_n(\eta_2 a)}{H^{(1)}_n(\eta_2 a)} \bigg),
	\end{split}
\end{equation}
	where
	\begin{equation}\label{m21}
		W_R= - \frac{n^2\beta^2}{a^2}\bigg(\frac{1}{\eta_2^2}-\frac{1}{\eta_1^2} \bigg)^2
		+\bigg(\frac{J'_n(\eta_1 a)}{\eta_1 J_n(\eta_1 a)}-\frac{H^{(1)\prime}_n(\eta_2 a)}{\eta_2 H^{(1)}_n(\eta_2 a)} \bigg)
		\bigg(\frac{k_1^2J'_n(\eta_1 a)}{\eta_1 J_n(\eta_1 a)}-\frac{k_2^2H^{(1)\prime}_n(\eta_2 a)}{\eta_2 H^{(1)}_n(\eta_2 a)} \bigg).
	\end{equation}	
	Note that $A_R$ and $C_R$ are even functions of $\beta$ and $n$, while $B_R$ and $D_R$ are odd functions of these variables.
	
	In the particular case where $\mathbf{R}=\mathbf{R}'$, Eq.~(\ref{m15}) reduces to
\begin{eqnarray}\label{m23}
	\mathbf{G}^{(\mathrm{sc})}(\mathbf{R},\mathbf{R},\omega)
	&=&\frac{i}{8\pi}\sum_{n=-\infty}^\infty\int_{-\infty}^{\infty} d\beta
	\bigg\{A_R\bigg[	
	\frac{n^2}{\eta_2^2r^2}H_n^{(1)\,2}(\eta_2 r)\hat{\mathbf{r}}\hat{\mathbf{r}}
	+ H_n^{(1)\prime \,2}(\eta_2 r)\hat{\boldsymbol{\varphi}}\hat{\boldsymbol{\varphi}}\bigg]
	\nonumber\\&&\mbox{}
	+ C_R\frac{\beta^2}{k_2^2}\bigg[
H_n^{(1)\prime \,2}(\eta_2 r)\hat{\mathbf{r}}\hat{\mathbf{r}}
+\frac{n^2}{\eta_2^2 r^2}H_n^{(1)\,2}(\eta_2 r)\hat{\boldsymbol{\varphi}}\hat{\boldsymbol{\varphi}}
+\frac{\eta_2^2}{\beta^2} H_n^{(1)\,2}(\eta_2 r)\hat{\mathbf{z}}\hat{\mathbf{z}}\bigg]	
\nonumber\\&&\mbox{}	
+(B_R+D_R)
\frac{n\beta}{\eta_2 k_2r}H_n^{(1)}(\eta_2 r)H_n^{(1)\prime}(\eta_2 r)
(\hat{\mathbf{r}}\hat{\mathbf{r}}
+\hat{\boldsymbol{\varphi}}\hat{\boldsymbol{\varphi}})\bigg\}.
\end{eqnarray}
It is clear from Eq.~(\ref{m23}) that the tensor $\mathbf{G}^{(\mathrm{sc})}(\mathbf{R},\mathbf{R},\omega)$ is diagonal in the cylindrical coordinate basis. Note that expression (\ref{m23}) contains not only the terms associated with the coefficients $A_R$ and $C_R$ but also the terms associated with the coefficients $B_R$ and $D_R$.
This feature is different from the corresponding result of Ref.~\cite{Ellingsen2010} for a cylindrical cavity.

From Eq.~(\ref{m23}), we find
\begin{eqnarray}\label{m24}
	\mathrm{Tr}[\mathbf{G}^{(\mathrm{sc})}(\mathbf{R},\mathbf{R},\omega)]
	&=&\frac{i}{8\pi}\sum_{n=-\infty}^\infty\int_{-\infty}^{\infty}d\beta
	\bigg\{\bigg(A_R+C_R\frac{\beta^2}{k_2^2}\bigg)\bigg[	
	\frac{n^2}{\eta_2^2 r^2}H_n^{(1)\,2}(\eta_2 r)
	+H_n^{(1)\prime \,2}(\eta_2 r)\bigg]
	+C_R\frac{\eta_2^2}{k_2^2}H_n^{(1)\,2}(\eta_2 r)
	\nonumber\\&&\mbox{}		
	+2(B_R+D_R)\frac{n\beta}{\eta_2 k_2r}H_n^{(1)}(\eta_2 r)H_n^{(1)\prime}(\eta_2 r)
	\bigg\}.
\end{eqnarray}
The spatial dependence of $\mathrm{Tr}[\mathbf{G}^{(\mathrm{sc})}(\mathbf{R},\mathbf{R},\omega)]$ is determined by the functions $H_n^{(1)}(\eta_2 r)$. 

For the imaginary frequency $\omega=iu$, we have
$k_j=i\kappa_j$ and $\eta_j=iq_j$, where $\kappa_j=\sqrt{\epsilon_j(iu)}\,u/c$ and $q_j=\sqrt{\kappa_j^2+\beta^2}=\sqrt{\epsilon_j(iu) u^2/c^2+\beta^2}$ for $j=1$ and 2.
Note that $\epsilon_j(iu)$ is real and positive and hence so is $q_j$.
Using the relations $J_n(ix)=i^n I_n(x)$ and $H_n^{(1)}(ix)=(2/\pi) i^{-n-1}K_n(x)$, we find
\begin{eqnarray}\label{m25}
	\mathrm{Tr}[\mathbf{G}^{(\mathrm{sc})}(\mathbf{R},\mathbf{R},iu)]
	&=&\frac{1}{4\pi^2}\sum_{n=-\infty}^\infty\int_{-\infty}^{\infty}d\beta
	\bigg\{\bigg(A-C\frac{\beta^2}{\kappa_2^2}\bigg)
	\bigg[\frac{n^2}{q_2^2 r^2}K_n^{2}(q_2 r)
	+K_n^{\prime \,2}(q_2 r)\bigg]
	-C\frac{q_2^2}{\kappa_2^2}K_n^{2}(q_2 r)
	\nonumber\\&&\mbox{}		
	-2i(B+D)\frac{n\beta}{q_2 \kappa_2r}K_n(q_2 r)K_n^{\prime}(q_2 r)
	\bigg\},
\end{eqnarray}	
where
\begin{equation}\label{m27}
	\begin{split}
		A&=\frac{1}{W}\frac{I_n(q_2a)}{K_n(q_2a)}\bigg[\frac{n^2\beta^2}{a^2} \bigg(\frac{1}{q_2^2}-\frac{1}{q_1^2} \bigg)^2
		+\bigg(\frac{I'_n(q_1 a)}{q_1 I_n(q_1 a)}-\frac{I'_n(q_2 a)}{q_2 I_n(q_2 a)} \bigg)
		\bigg(\frac{\kappa_1^2I'_n(q_1 a)}{q_1 I_n(q_1 a)}-\frac{\kappa_2^2K^{\prime}_n(q_2 a)}{q_2 K_n(q_2 a)} \bigg)\bigg],\\
		C&=\frac{1}{W}\frac{I_n(q_2a)}{K_n(q_2a)}\bigg[\frac{n^2\beta^2}{a^2} \bigg(\frac{1}{q_2^2}-\frac{1}{q_1^2} \bigg)^2
		+\bigg(\frac{I'_n(q_1 a)}{q_1 I_n(q_1 a)}-\frac{K^{\prime}_n(q_2 a)}{q_2 K_n(q_2 a)} \bigg)
		\bigg(\frac{\kappa_1^2I'_n(q_1 a)}{q_1 I_n(q_1 a)}-\frac{\kappa_2^2I'_n(q_2 a)}{q_2 I_n(q_2 a)} \bigg)\bigg],\\
	B&=D=\frac{i}{W}\frac{I_n(q_2a)}{K_n(q_2a)}\frac{\kappa_2}{q_2}\frac{n\beta}{a} 
	\bigg(\frac{1}{q_2^2}-\frac{1}{q_1^2} \bigg)\bigg(\frac{I'_n(q_2 a)}{I_n(q_2 a)}-\frac{K^{\prime}_n(q_2 a)}{K_n(q_2 a)}\bigg),			
	\end{split}
\end{equation}
with
\begin{equation}\label{m26}
	W=\frac{n^2\beta^2}{a^2}\bigg(\frac{1}{q_2^2}-\frac{1}{q_1^2} \bigg)^2
	+\bigg(\frac{I'_n(q_1 a)}{q_1 I_n(q_1 a)}-\frac{K^{\prime}_n(q_2 a)}{q_2 K_n(q_2 a)} \bigg)
	\bigg(\frac{\kappa_1^2I'_n(q_1 a)}{q_1 I_n(q_1 a)}-\frac{\kappa_2^2K^{\prime}_n(q_2 a)}{q_2 K_n(q_2 a)} \bigg).
\end{equation}
Note that $\mathrm{Tr}[\mathbf{G}^{(\mathrm{sc})}(\mathbf{R},\mathbf{R},iu)]$ is real.
The spatial dependence of $\mathrm{Tr}[\mathbf{G}^{(\mathrm{sc})}(\mathbf{R},\mathbf{R},iu)]$ is determined by the monotonic functions $K_n(q_2r)$. 
	
\end{widetext}

\section{Dawson-function model for the silica dielectric permittivity dispersion}
\label{sec:Dawson}

In order to describe the frequency dispersion of the dielectric permittivity $\epsilon(\omega) = \epsilon^\prime(\omega) + i \epsilon^{\prime\prime}(\omega)$ of silica for the wavelength in the range from 7 to 50 $\mu$m, we employ the Dawson-function model suggested by Meneses and co-authors in Ref.~\cite{Meneses}.
According to this model, we have
\begin{eqnarray}\label{b1}
	\epsilon^{\prime}(\eta) &=& \epsilon_{\infty} + 2 \sum\limits_{j} \frac{\alpha_j}{\sqrt{\pi}} \Big( D(2 \sqrt{\ln{2}}\; (\eta + \eta_{j})/\sigma_j )\nonumber\\&&\mbox{} 
	-D(2 \sqrt{\ln{2}}\; (\eta - \eta_{j})/\sigma_j ) \Big),\nonumber\\
	\epsilon^{\prime \prime} (\eta) &=&  \sum\limits_{j} \alpha_j\Big( e^{-4 \ln{2}\; (\eta - \eta_{j})^2/\sigma_j^2} 
	- e^{-4 \ln{2}\; (\eta + \eta_{j})^2/\sigma_j^2}\Big).
	\nonumber\\
\end{eqnarray}
Here, $\eta$ is the frequency measured in $\text{cm}^{-1}$, $\epsilon_{\infty} = 2.1232$ is the permittivity of silica at high frequencies, $D(x) = e^{-x^2} \int\limits_{0}^{x} e^{t^2} dt$ is the Dawson function, and $\alpha_j$, $\eta_{j}$, and $\sigma_j$ are the model fitting parameters that were suggested in Ref.~\cite{Kitamura} for silica and are presented in Table \ref{table}.

\begin{table}[b]
	\caption{\label{table} Model parameters proposed in Ref.~\cite{Kitamura} and used to fit different experimental data on silica glass permittivity.}
	\begin{ruledtabular}
		\begin{tabular}{llll}			
$j$ & $\alpha_j$ & $\eta_{j}$ (cm$^{-1}$) & $\sigma_j$ (cm$^{-1}$) \\						
			\colrule
1 & 3.7998 & 1089.7  & 31.454 \\
2 & 0.46089 & 1187.7 & 100.46 \\
3 & 1.2520 & 797.78 & 91.601  \\
4 & 7.8147 & 1058.2 & 63.153 \\
5 & 1.0313 & 446.13 & 275.111 \\
6 & 5.3757 & 443.00 & 45.220 \\ 
7 & 6.3305 & 465.80 & 22.680 \\
8 & 1.2948 & 1026.7 & 232.14 \\
		\end{tabular}
	\end{ruledtabular}
\end{table}

Note that the Dawson function $D(z)$ can be generalized for the entire complex plane of its argument $z$ and can be written as
\begin{equation}\label{b2}
	D(z) = \dfrac{i \sqrt{\pi}}{2} \left( e^{-z^2} - w(z) \right),
\end{equation}
where \begin{equation}\label{b3}
	w(z) = e^{-z^2} \bigg( 1 + \dfrac{2 i}{\sqrt{\pi}} \int\limits_{0}^{z} e^{t^2} dt \bigg)
\end{equation}
is the Faddeeva function. 
By making use of Eq.~(\ref{b2}), one can describe the model (\ref{b1}) in a rather elegant form
\begin{eqnarray}
	\epsilon(\eta) &=& \epsilon_{\infty} + \sum\limits_{j} i \alpha_j \left( w(z_{j,-}) - w(z_{j,+}) \right), \nonumber\\
	z_{j,\pm} &=& \dfrac{ 2 \sqrt{\ln{2}}\; (\eta \pm \eta_{j}) }{\sigma_j}.
\end{eqnarray}

When dealing with the nonresonant part of the Casimir-Polder potential, one has to compute the dielectric permittivity for a purely imaginary frequency. In this case, owing to the property $w(x+iy) = w^*(-x+iy)$ (for real $x$ and $y$), one can obtain
\begin{equation}
	\epsilon(i \eta) = \epsilon_{\infty} + 2 \sum\limits_{j} \alpha_j \text{Im}\left[ w(\zeta_{j}) \right],
\end{equation}
where $\zeta_{j} =  2 \sqrt{\ln{2}}\; (i\eta + \eta_{j}) / \sigma_j$.

\end{document}